\documentclass[aip,rsi,amssymb,reprint,graphicx]{revtex4-2}
\usepackage{graphicx}
\usepackage[]{hyperref}
\hypersetup{colorlinks=true,linkcolor=blue,citecolor=blue,urlcolor=blue,pdfpagemode=UseNone}

\begin{document}

\title{Measurements of nematic susceptibility with phase sensitive nuclear magnetic resonance in pulsed strain fields}

\author{C. Chaffey}
\affiliation{Department of Physics and Astronomy, University of California Davis, Davis, CA }
\author{C. Williams}
\affiliation{Department of Physics and Astronomy, University of California Davis, Davis, CA }
\author{M. A. Tanatar}

\author{S. L. Bud'ko}

\author{P. C. Canfield}

\affiliation{Ames Laboratory U.S. DOE and Department of Physics and Astronomy,Iowa
State University, Ames, Iowa 50011, USA}
\author{N. J. Curro}
\email{njcurro@ucdavis.edu}
\affiliation{Department of Physics and Astronomy, University of California Davis, Davis, CA }

\date{\today}

\begin{abstract}

{We present nuclear magnetic resonance data in BaFe$_2$As$_2$ in the presence of pulsed strain fields that are interleaved in time with the radiofrequency excitation pulses.  In this approach, the precessing nuclear magnetization acquires a phase shift that is proportional to the strain and pulse time.  The sensitivity in this approach is limited by the homogeneous decoherence time, $T_2$, rather than the inhomogeneous linewidth. We measure the nematic susceptibility as a function of temperature, and demonstrate three orders of magnitude improvement in sensitivity.  This approach will enable studies of the strain response in a broad range of materials that previously were inaccessible due to inhomogeneous broadening.}

\end{abstract}
\maketitle

\section{Introduction}

In recent years there has been growing attention to the role of electronic nematicity in strongly correlated electron materials, where the electronic degrees of freedom spontaneously break an underlying discrete symmetry of the lattice \cite{FradkinNematicReview}.  This phenomenon is exemplified in the parent compounds of many iron-based superconductors, where the the Fe $3d$ orbitals undergo  ferro-orbital ordering, accompanied by a tetragonal to orthorhombic structural transition and often the onset of long-range antiferromagnetic order  \cite{doping122review,FernandesNematicPnictides}. Below this temperature in the nematic, or orthorhombic, phase, the Fe $d_{xz}$ and $d_{yz}$ orbitals become nondegenerate, with an energy splitting
on the order of 40 meV, and different occupation levels \cite{NaFeAsARPES2012}. Doping tends to suppress the long-range nematic order and stabilizes unconventional superconductivity. Elastoresistance measurements of the nematic susceptibility \cite{FisherScienceNematic2012,Kuo2015,Worasaran2021} revealed the presence of a nematic quantum phase critical point that may be responsible for driving the superconductivity in this system \cite{FernandesSchmalianNatPhys2014,KivelsonNematicQCP2015,ScalapinoNematicQCP2015,Lederer2017}.

Several techniques have been developed to probe the nematic degrees of freedom. Anisotropic resistivity \cite{TanatarDetwin,IronArsenideDetwinnedFisherScience2010}, elastoresistance \cite{FisherScienceNematic2012}, elastic constants \cite{NematicElasticPRL2010,IshidaFeSePRL2015,MeingastBaFe2As2strain2016,MeingastNematicSusceptibility2016}, thermopower \cite{ThermopowerNematicityPRL}, electronic Raman scattering \cite{Ba122RamanPRL2013}, polarized light image color analysis \cite{TanatarTensileStressPRB,TanatarPRL2016} and optical conductivity \cite{Mirri2015} probe bulk anisotropies.
Nuclear magnetic resonance (NMR) and neutron scattering have been  utilized to investigate the effect of nematicity on the spin fluctuations \cite{PhysRevB.89.214511,Kissikov2018,DaiRMP2015,StrainBa122neutronsPRB2015,StrainedPnictidesNS2014science}. 
NMR studies of quadrupolar nuclei are sensitive to charge degrees of freedom, and have also been utilized to study both nematic fluctuations \cite{DioguardiPdoped2015}, and changes to the Fe $3d$ orbital occupations in response to external strain fields \cite{EFGnematicity}.  The latter provides a direct microscopic measure of the nematic susceptibility, and has the advantage that it can be measured in the superconducting state, where other measurements such as elastoresistance are unable to operate.

NMR measurements of the nematic susceptibility probe the response of the NMR resonance frequencies as a function of applied strain.  Strain can couple to the nuclear spins via either the Knight shift or the electric field gradient (EFG) tensors. These quantities inherit the point group symmetry of the lattice, and if this symmetry is lowered by strain, then these tensors can acquire new asymmetries or off-diagonal components.  Note that the Knight shift tensor depends on both the hyperfine coupling and the electronic spin susceptibility, and both could change in response to strain. In the case of FeSe only the latter appears to become anisotropic in the nematic phase \cite{Baek2015}. In BaFe$_2$As$_2$, evidence to date says that the Knight shift tensor does not respond to strain, although the EFG tensor changes dramatically.  The NMR resonance frequencies depend sensitively on these parameters, and thus one can probe the nematic susceptibility of a material by observing the linear response of the resonance frequency to strain. 

A straightforward approach to measuring the nematic suscepibility with NMR is to observe a shift in the spectrum while applying a static strain field.  However in order to detect a response, the frequency shift must be on the order of the spectral linewidth, $(T_{2}^*)^{-1}$, which often is inhomogeneously broadened.   This quantity can exceed the intrinsic decoherence rate, $T_2^{-1}$, by orders of magnitude, especially in the presence of doping which creates an inhomogeneous distribution of local EFGs.  In such cases the sensitivity of the susceptibility measurements can be severely limited.  On the other hand, if the strain field is pulsed while the nuclear spins are coherently precessing, then the shift of the resonance frequency will alter the phase of the precessing magnetization and can be observed via quadrature detection.  This process forms the basis of ``spin-warp" imaging widely used in magnetic resonance imaging, in which a magnetic field gradient is pulsed during the evolution times of a spin echo sequence  \cite{Edelstein1980,Johnson1983}. In this case it is $T_2^{-1}$, rather than the inhomogeneous linewidth, that limits the sensitivity of the  response measurement. A pulsed-strain approach should thus enable measurements in systems with inhomogeneous broadening or manifesting smaller intrinsic suscepetibilities. Indeed, strain pulses were used to study the strain response of color centers in silicon carbide using optically detected magnetic resonance \cite{Falk2014}. 

The approach outlined here is similar to AC strain measurements that have been utilized to measure elastoresistance and elastocaloric effects \cite{Hristov2018,Ikeda2019}, but an important difference is that the strain fields we apply are pulsed rather than  sinusoidal.  In this case the response function is a convolution of the frequency dependent nematic susceptibility with the square-wave driving voltage, giving rise to an exponential rise or decay of the nematicity with time constant $\tau_{nem}$.  This time scale is determined by the electronic degrees of freedom, thus $\tau_{nem}\sim 10^{-9}$ s.  Another important time scale, $\tau_{s}$, is determined by the speed of sound as the strain pulse traverses the length of the sample, which we estimate to be $\sim 10^{-7}$ s. On the other hand, the shortest time scale in our measurements is at least a microsecond, which is larger than either $\tau_{nem}$ or $\tau_{s}$, therefore  we are operating in the quasi-static limit.

\section{Experiments}

In order to examine the feasibility of a pulsed-strain approach to measuring the nematic susceptibility, we studied the response in a well-characterized system, BaFe$_2$As$_2$ \cite{doping122review}. A single crystal grown by self-flux was cut to dimensions $\sim 1.6$ mm long by $\sim 0.3$ mm wide, with the long axis parallel to the (110) direction (in the tetragonal basis), and mounted to a commercial piezoelectric strain cell (CS100, Razorbill Instruments) \cite{Hicks2014}. The crystal was mounted using epoxy (UHU plus endfest 300) in order to orient the magnetic field in-plane, as shown in Fig. \ref{fig:circuit} and described in \cite{KissikovStrainProbe}. Uniaxial stress was applied along the long-axis by piezoelectric stacks giving rise to a strain $\varepsilon= (\varepsilon_{xx} - \varepsilon_{yy})/2$, with $B_{2g}$ symmetry. Because the  Poisson ratio is non-zero, there will also be strain fields with $A_{1g}$ symmetry, however the nematic order in this material couples primarily to the $B_{2g}$ channel and will dominate the response \cite{Curro2022,Kuo2015}. The static sample length displacement, $\Delta L$ was measured by a capacitive dilatometer, and strain, $\varepsilon = \Delta L/L_0$, was determined based an unstressed length of $L_0 = 0.55$ mm at room temperature.

\subsection{Pulsed Strain}

\begin{figure}[!h]
    \includegraphics[width=\linewidth]{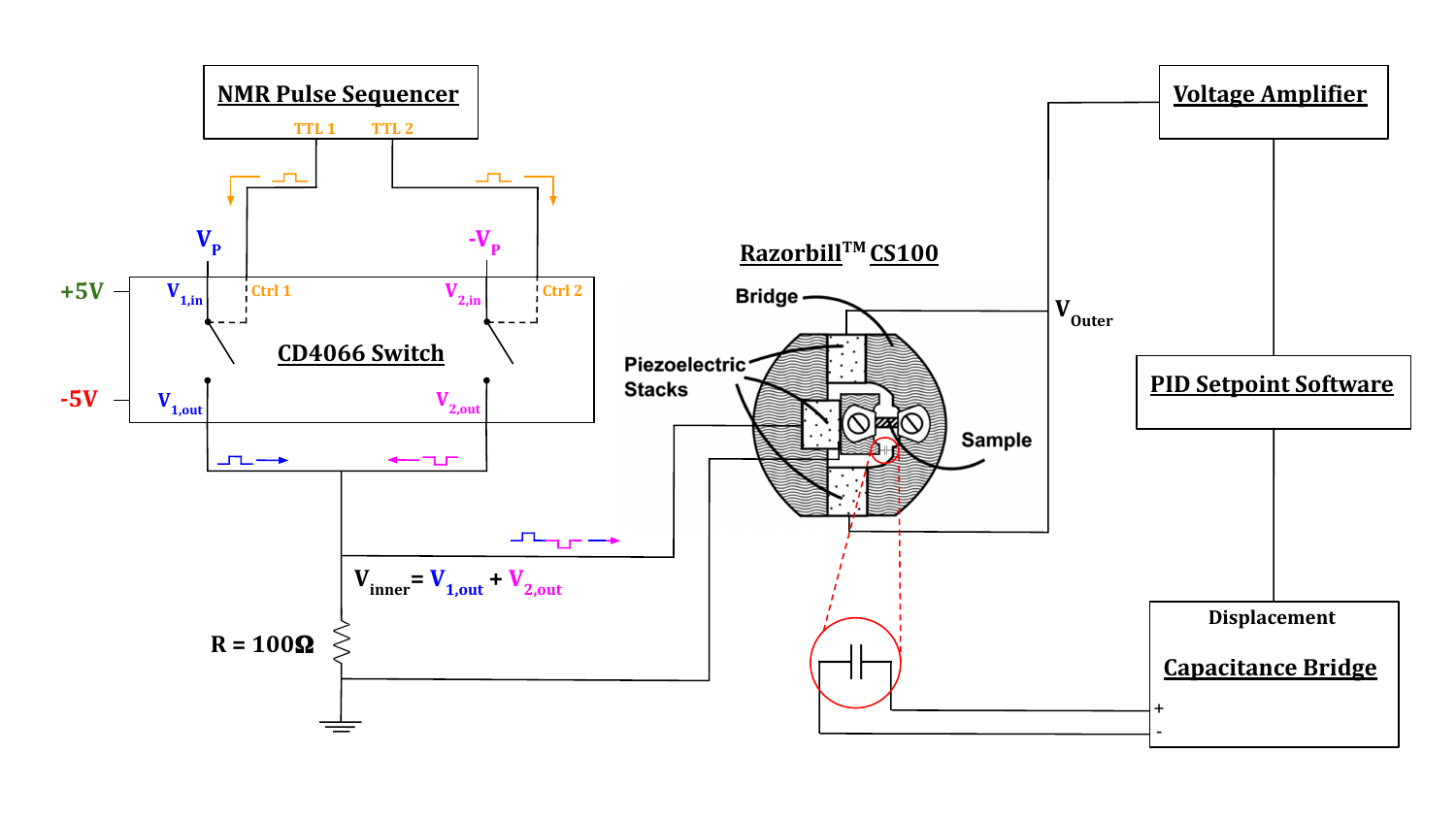}
    \includegraphics[width=\linewidth]{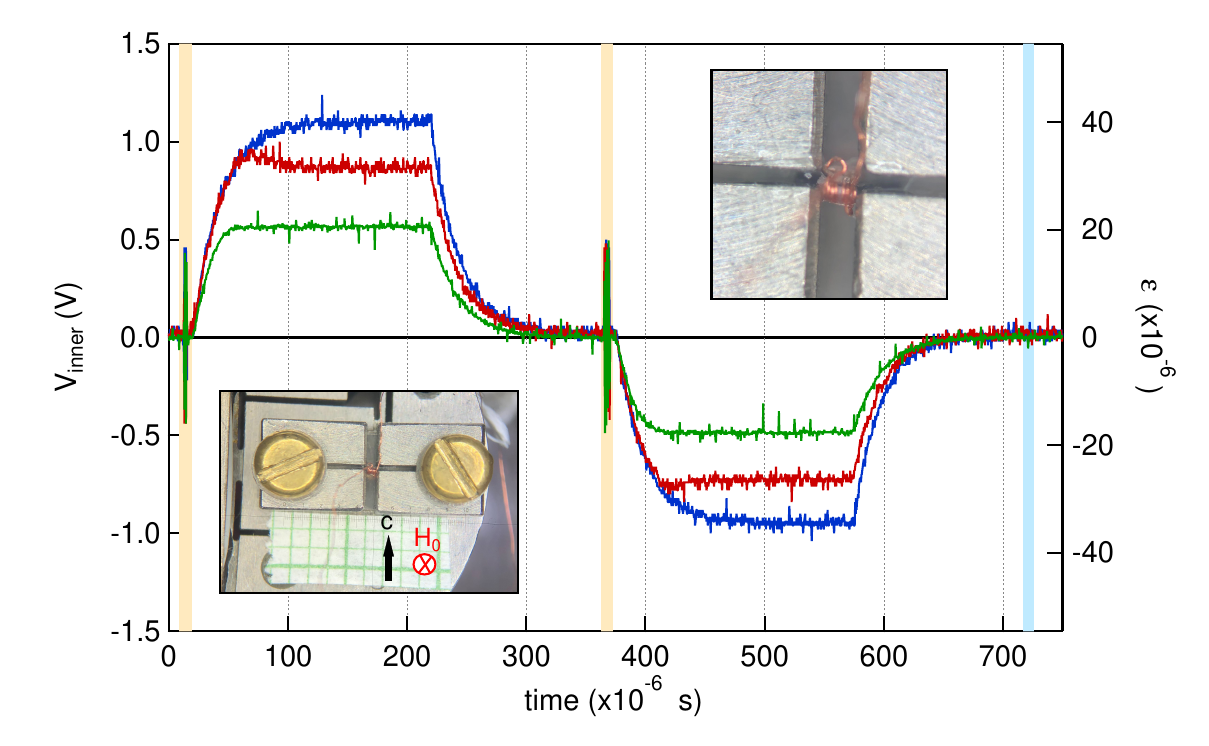}
    \caption{\label{fig:circuit}  (Upper panel) Circuit diagram illustrating the creating of the strain pulses.  TTL pulses control analog switches  with DC voltages set to $\pm V_P$, which control the inner stack of the CS100 strain device.  The outer stack is held at constant displacement using the PID feedback control and the capacitive dilatometer. (Lower panel) Measured voltage across the inner stack as a function of time for three different values of $V_P$ (corresponding strain values shown on the right side axis). The light orange vertical bands indicate the NMR radiofrequency pulses, and the light blue vertical band indicates the position of the echo. The lower inset shows the crystal mounted in the device with the surrounding coil, and the direction of the applied field. The upper insert shows a blow up of the mounted crystal. }
\end{figure}

\begin{figure}[!h]
    \includegraphics[width=\linewidth]{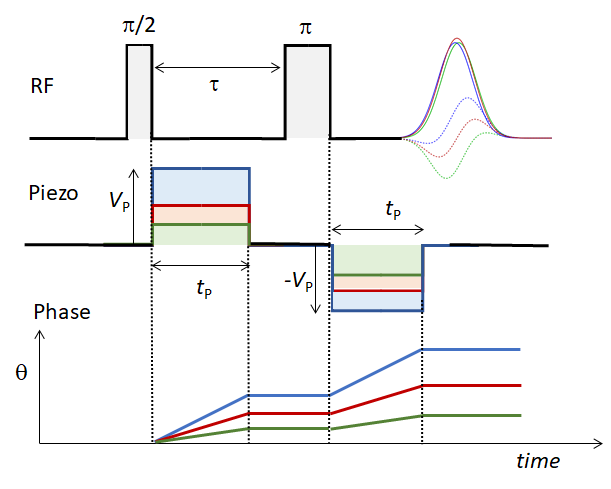}
    \caption{\label{fig:timing}  Strain pulses are applied during the times of free precession, between the two radiofrequency pulses.  The phase of the echo in the rotating frame will change by an amount that is proportional to the strain pulse width, $t_p$, times the strain pulse magnitude, $V_p$. $t_P$ was varied between 200 and 670 $\mu$s.  The pulse length for a 90 degree pulse was 2.5 $\mu$s, and $V_P$ varied from -2 to +2 V.  }
\end{figure}

Th strain device utilizes two sets of piezos, the inner and the outer stacks, as illustrated in Fig. \ref{fig:circuit}. In order to implement the strain pulses,  we used the outer stack to hold a static strain field, while applying voltage pulses  to the inner stack.  These strain pulses were created by utilizing TTL pulses from an NMR spectrometer to drive analog switches (CD4066BE, Texas Instruments) that were timed to turn on immediately following the $\pi/2$ and $\pi$ radiofrequency pulses, as illustrated as in Fig. \ref{fig:timing}. The time dependent voltage across the inner stacks is shown in the lower panel of Fig. \ref{fig:circuit}. The resistance $R=100$ $\Omega$ shown in Fig. \ref{fig:circuit} is in parallel with the piezo stack, which has an intrinsic capacitance ($\sim 0.2$ $\mu$F) and was chosen to provide a rise time 
that is shorter than $T_2$, so that the strain pulse can be applied between the  NMR radiofrequency pulses. 

\begin{figure}[!h]
    \includegraphics[width=\linewidth]{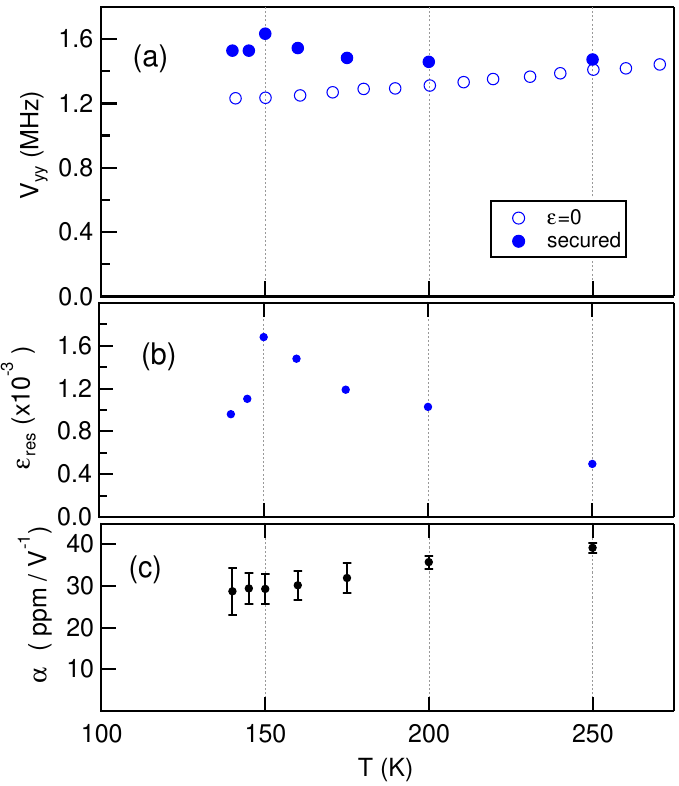}
    \caption{\label{fig:calibrations}  (a) Quadrupolar splitting $\nu_{yy} = |f_+ - f_-|/2$ as a function of temperature, compared with that in a free-standing sample (from \cite{takigawa2008}.) (b) Residual strain versus temperature, based on $V_{yy}$ and using the nematic susceptibility reported in \cite{EFGnematicity}. (c) The strain calibration factor, $\alpha = d\varepsilon/dV$, versus temperature as discussed in the text.  }
\end{figure}

Static displacements are measured via a capacitance bridge, however the time constant for the bridge precludes any measurements during a short strain pulse.  We estimate the strain during the pulses using a temperature dependent calibration factor, $\alpha = \varepsilon/V_P$, shown in Fig. \ref{fig:calibrations}(c), which was determined by measurements of the displacement during static applied voltage levels. This approach was utilized in previous AC strain measurements \cite{Hristov2018,Ikeda2019}, and  can lead to overestimates of the strain applied due to the frequency dependence of the calibration factor.   As discussed below, the NMR response with pulsed and static strain give similar values, suggesting that the calibration approach is reasonable.
The dynamic strain can also give rise to DC heating effects of the strain device on the order of a few Kelvin at finite frequencies, depending on the magnitude of the driving voltage, because the large capacitances and low thermal conductivity of the piezo stacks  \cite{Hristov2018}. To minimize such effects we operated with low driving voltages ($\leq2$ V).  

A potential issue with the application of voltage pulses to the strain cell is that  natural vibration resonances could be excited in the device.  We observed resonances in the range of  $\sim 2-30$ kHz for our device suggesting that time scales faster than 1 ms could be problematic. In practice, however, the low values of $V_P$ we utilized did not appear to introduce any oscillations measured in $V_{inner}$, as shown in Fig. \ref{fig:circuit}. We speculate that low mass of the strain cell ensures that these resonances lie outside the bandwidths of our pulses.

\subsection{NMR phase sensing}

The goal of the experiments is to probe the strain response of the spin $I=3/2$ $^{75}$As nuclei in this material.  The resonance frequencies are determined by the eigenvalues of the nuclear spin Hamiltonian:
\begin{equation}
  \mathcal{H} = \gamma\hbar{\mathbf{I}}\cdot(1+\mathbf{K})\cdot \mathbf{H}_0 + \frac{h\nu_{zz}}{6}[3{I}_z^2-\hat{I}^2 + \eta\left(\hat{I}_x^2-{I}_y^2\right)]
  \label{eqn:Hnuc}
\end{equation}
where $\gamma=0.7292$ kHz/G is the gyromagnetic ratio,  ${I}_{\alpha}$ are the nuclear spin operators, $\mathbf{K}$ is the Knight shift tensor, $\nu_{zz}$ is the largest eigenvalue of the electric field gradient (EFG) tensor, and $\eta = (\nu_{yy} -\nu_{xx})/(\nu_{xx} + \nu_{yy})$ is the asymmetry parameter.   To second order in perturbation theory, the resonance frequencies of the satellites and central transitions for $c\perp \mathbf{H}_0$ are given by:
\begin{eqnarray}
  f_{\pm} &=& \gamma H_0(1 + K_{\perp} + \Delta K) \mp \frac{\eta + 1}{2}\nu_{zz} \\
\nonumber  f_{0} &=& \gamma H_0(1 + K_{\perp} + \Delta K) + \frac{(\eta-3)^2\nu_{zz}^2}{48\gamma H_0}
\label{eqn:freqs}
\end{eqnarray}
where $K_{\perp} = (K_{xx} + K_{yy})/2$ and $\Delta K = K_{yy} - K_{xx}$. If any of the three parameters $\eta$, $\nu_{zz}$ or $\Delta K$ changes in response to strain, then there will be shifts in $f_0$ and $f_{\pm}$.

\begin{figure*}
\centering
\begin{minipage}{.7\textwidth}
  \centering
  \includegraphics[width=\linewidth]{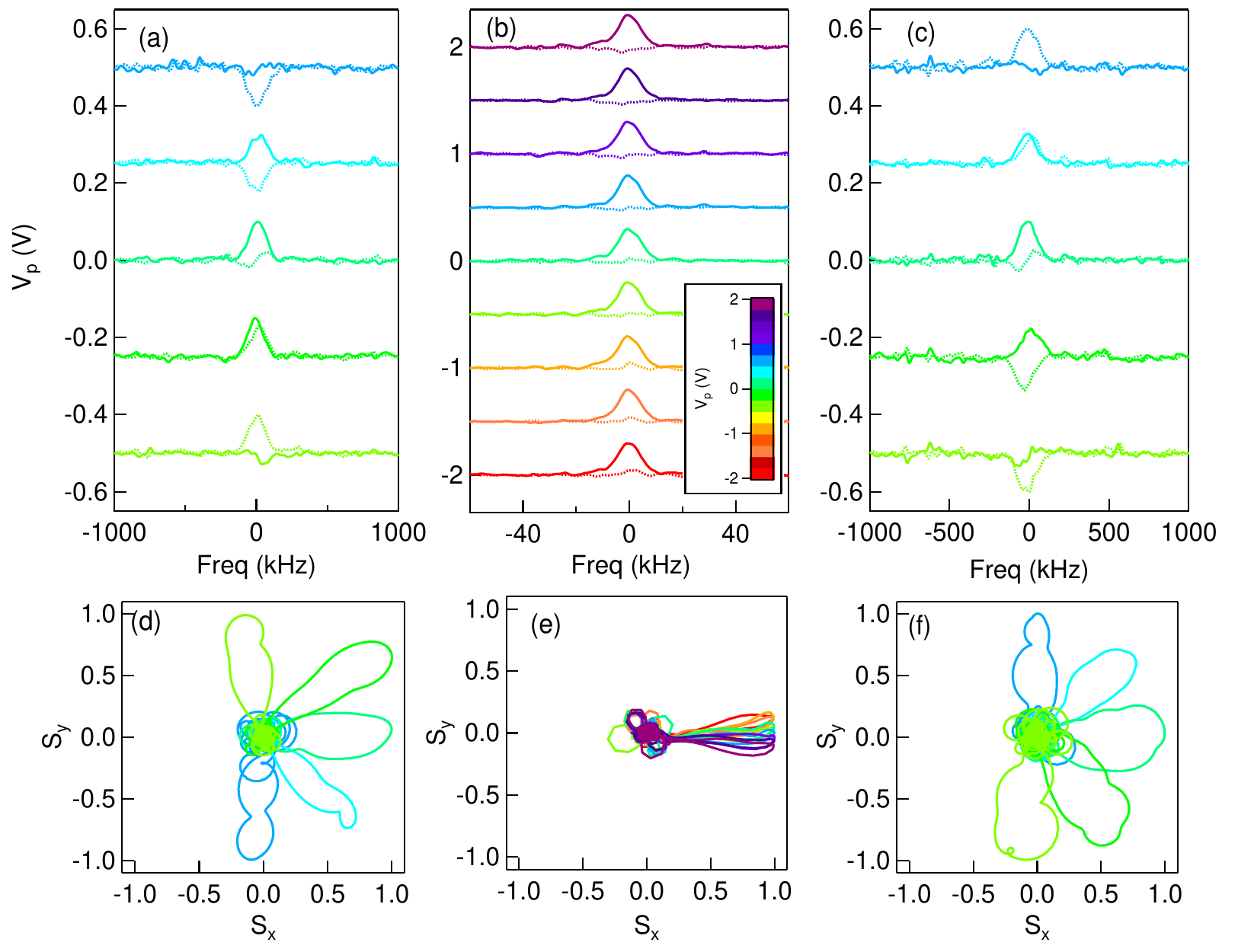}
\end{minipage}%
\begin{minipage}{.3\textwidth}
  \centering
  \includegraphics[width=\linewidth]{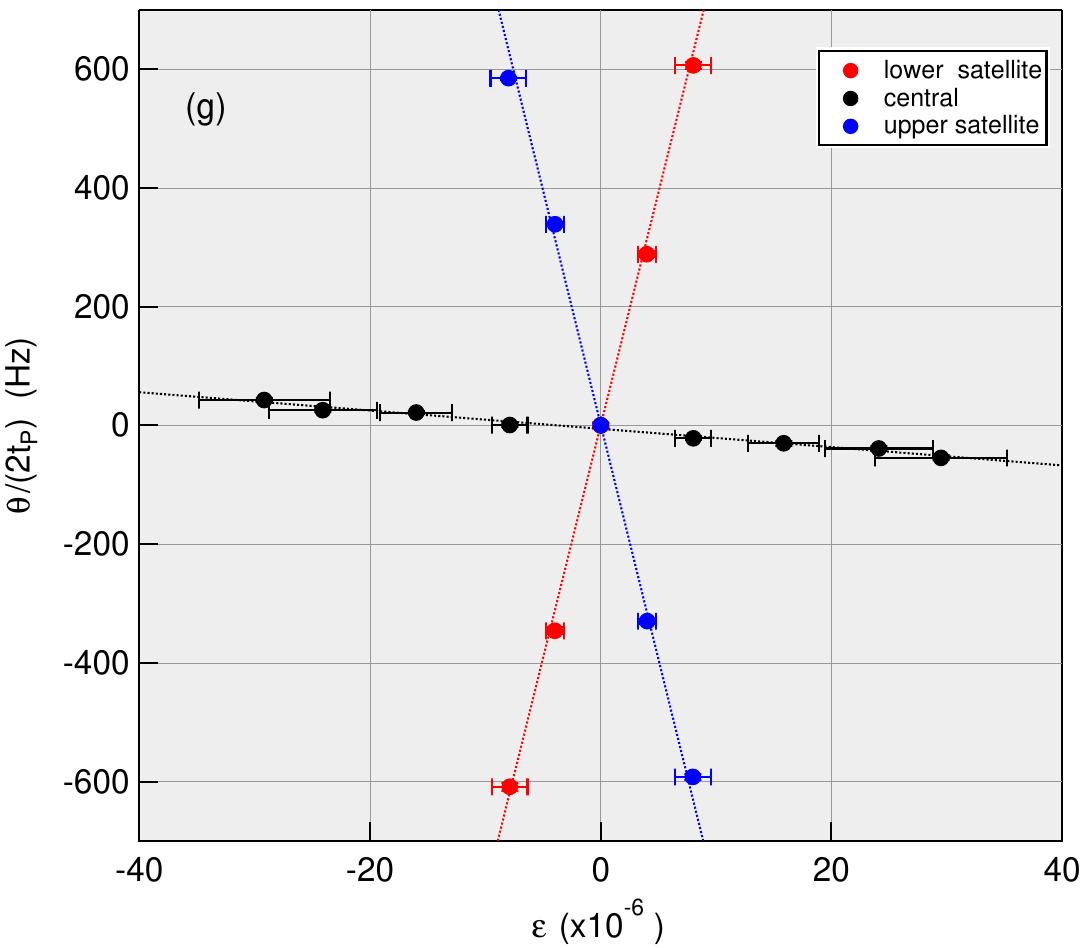}
\end{minipage}
\caption{\label{fig:rotations}  Normalized spectra versus frequency for the (a) upper, (b) central and (c) lower transitions of the $^{75}$As. Solid and dotted lines correspond to the projection along the $x$ and $y$ axes in the rotating frame. The frequency axes have been offset by the corresponding resonance frequencies, and the spectra have been offset vertically by the value of $V_{p}$.  Plots (d), (e) and (f) show the same data in which $S_y$ is plotted versus $S_x$, with frequency as an implicit parameter. The color scale corresponds to $V_p$ and is the same for all plots. (g) Echo phase over strain pulse time (in cycles per second) versus applied strain at 140 K.  The dotted lines are linear fits using least orthogonal distance method, with slopes $\partial f_-/\partial \varepsilon = 78\pm 8$ MHz/strain, $\partial f_0/\partial \varepsilon = -1.5\pm 0.1$ MHz/strain, and $\partial f_+/\partial \varepsilon = -79\pm 8$ MHz/strain.  }
\end{figure*}

For the pulse sequence shown in Fig. \ref{fig:timing}, the $\pi/2$ pulse creates a magnetization that precesses in the plane perpendicular to $\mathbf{H}_0$ at the resonance frequency of the particular transition. In the absence of any perturbing strain pulses, the magnetization would remain static in the rotating frame during the evolution time, $\tau$, between the $\pi/2$ and $\pi$ pulses. However, if a strain pulse during this period shifts the resonance frequency by an amount, $\Delta f$, then the magnetization accumulates an extra phase that is proportional to the product of $\Delta f$ and $t_P$, the duration of the strain pulse. The second strain pulse is inverted so that phase accumulation continues even though the direction of precession in the rotating frame is reversed after the $\pi$ pulse.  The end result is that the echo acquires a net phase $\theta = 2\Delta f t_p$.  This phenomenon is illustrated in Fig. \ref{fig:rotations}, where spectra are shown for both quadrature channels, $S_x$ and $S_y$, for all three transitions of the As.  The phase rotation is evident in the changes of the relative intensities of $S_x$ and $S_y$ components as a function of frequency as the magnitude of the piezo voltage, $V_p$, changes, as well as in plots of $S_x$ versus $S_y$ in the lower panels.  The effect is greatest for the two quadrupolar satellites, and smaller for the central transition.  Moreover, the lower satellite rotates counterclockwise ($\theta$ increases with $V_p$), whereas the upper and central transition rotate clockwise ($\theta$ decreases with $V_p$).  This behavior is summarized in Fig. \ref{fig:rotations}(g), which shows  $\Delta f = \theta/(2t_p)$ as a function of $V_p$.  Here we determine the angle $\theta = \arctan(I_y/I_x)$ (in cycles) where $I_{x,y}$ are the integrated intensities of the spectra $S_{x,y}$. 

We confirmed the behavior of the echo in response to strain pulses via numerical simulations of the full time-dependent Schrodinger equation using the QuTIP package \cite{Johansson2012,Johansson2013} with realistic pulse widths similar to those used in experiment, and modeling strain as a time dependent $\eta$.  The simulated echo phase exhibit behavior nearly identical to the experimental observations, and justifies the use of perturbation theory to express the frequencies of the transitions in Eq. \ref{eqn:freqs}.

\subsection{Susceptibility}

There are three distinct response functions that can give rise to the behavior in Fig. \ref{fig:rotations}: $\chi_{\eta} = \partial \eta/\partial \varepsilon$, $\chi_{K} = \partial \Delta K/\partial \varepsilon$, and $\chi_{\nu} = \partial \nu_{zz}/\partial \varepsilon$.  The linear response of the resonance frequencies are then given by ${\partial f_i}/{\partial \varepsilon} = \sum_j M_{ij}\chi_j$, where  $M_{ij} = \frac{\partial f_i}{\partial x_j}$ evaluated at zero strain and $x_{j} = \eta$, $\Delta K$ or $\nu_{zz}$. By measuring  $f_0$ and $f_{\pm}$, we can disentangle the strain susceptibility of each parameter individually: $\chi_j  = \sum_j{M}^{-1}_{ij} {\partial f_i}/{\partial \varepsilon}$. Using the fitted slopes in  Fig. \ref{fig:rotations}(g), we find 
\begin{eqnarray*}
    \chi_{\eta} &=& 92\pm 2/\textrm{strain}\\
    \chi_{\nu} &=& -54\pm 1 \textrm{MHz/strain}\\
    \chi_K&=& -0.336\pm 0.004\textrm{\%/strain}
\end{eqnarray*}
at 140 K.  The value of $\chi_{\eta}$ is comparable to that reported previously \cite{Kissikov2018}, however no change to the Knight shift has been previously observed.  The value we observe for $\chi_K$ is too small to detect via static strain techniques: for a maximum static strain $\varepsilon=10^{-3}$, $\Delta K$ corresponds to a shift of $\sim 0.5$ kHz, which is below the static linewidth of $\sim 1.2$ kHz.

\begin{figure}[!h]
    \includegraphics[width=\linewidth]{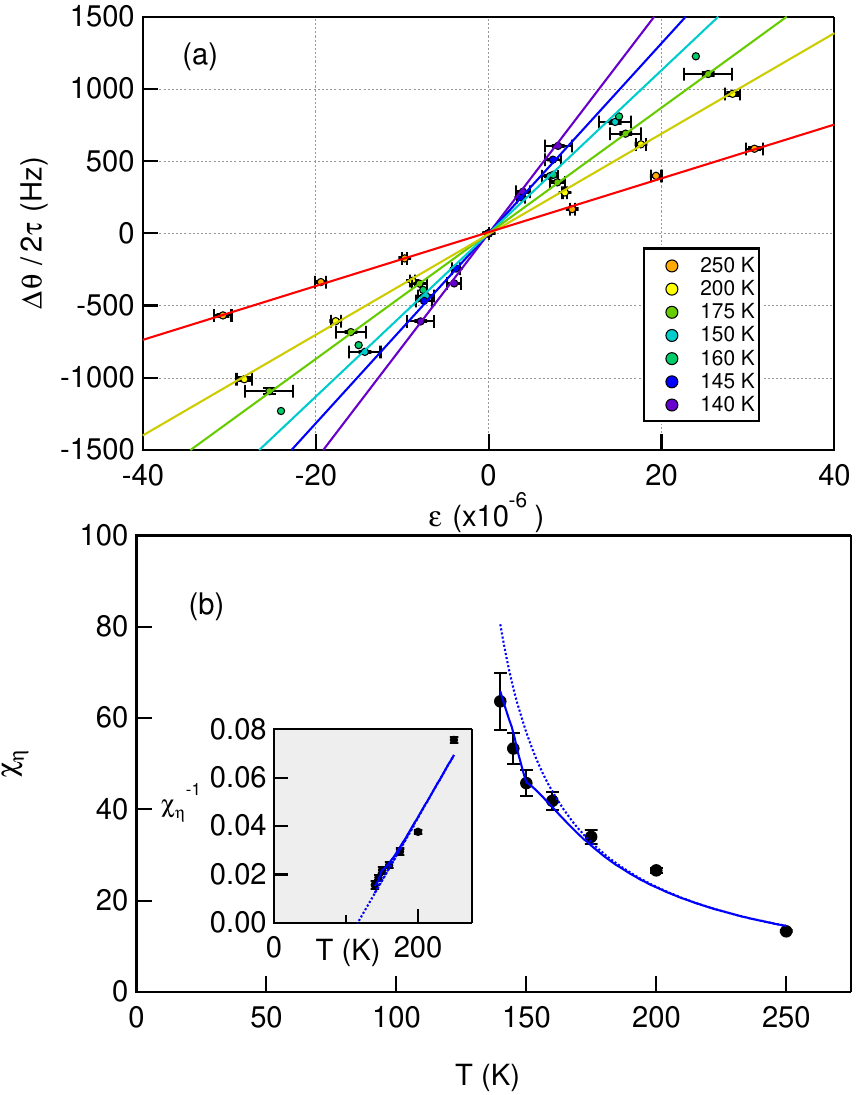}
    \caption{\label{fig:chi}  (a) Frequency shift of the lower satellite versus strain for several different temperatures. Solid lines are linear fits to the data. (b) Nematic susceptibility, $\chi_{\eta}$, as a function of temperature. The solid and dashed lines are fits with and without the residual strain. The inset shows the temperature dependence of the inverse of the susceptibility, as well as the corresponding fits.}
\end{figure}

The nonzero response $\chi_{\nu}$ is surprising, because $\nu_{zz}$ should only couple quadratically to strain with $B_{2g}$ symmetry.  We speculate that the linear response we observe reflects a finite $\varepsilon_{zz}$ strain field that is induced by the applied stress because of the finite Poisson ratio of our crystal. It is reasonable that $\chi_{\nu}< 0$ because $\nu_{zz}$ decreases with increasing $c$-axis length in the AFe$_2$As$_2$ (A = Ca, Ba, Sr) family \cite{Ca122EFGstudy}. 

Note that $\chi_{\eta}$ is the dominant source of the shift of the satellite resonance frequencies (Eq. \ref{eqn:freqs}).  Therefore it is possible to approximate $\chi_{\eta}\approx 2(\partial f_-/\partial \varepsilon)/\nu_{zz}(0)$, rather than measure the response of all three transitions at each temperature.  Fig. \ref{fig:chi}(a) displays the response of the lower transition for a series of temperatures, and Fig. \ref{fig:chi}(b) displays the temperature dependence of $\chi_{\eta}$.

For these experiments the sample displacement was held constant as the temperature was lowered.  A consequence of this condition is that thermal contraction of the crystal gives rise to a positive residual strain.  We estimated this residual strain by observing the EFG splitting of the satellite resonances in the absence of strain pulses. As shown in Fig. \ref{fig:calibrations}(a), the EFG splitting, $\nu_{yy}$, is temperature dependent and larger than in an the unstrained case. The residual strain can be determined using the nematic susceptibility measured previously using static strain \cite{EFGnematicity}: $\varepsilon_{res}=(2|\nu_{yy}|/\nu_{zz} - 1)/(\chi_0+C/(T-T_0))$, and is shown in Fig. \ref{fig:calibrations}(b) as a function of temperature using parameters reported previously.  The peak at 150 K is due to a software error, in which PID control was temporarily lost. 

The residual strain field will tend to suppress the true nematic susceptibility and can alter the temperature dependence of $\chi_{\eta}$.  To model this behavior, we 
assume that the nematic order parameter, $\phi$, can be described by a Ginzburg-Landau theory, and find the mean-field solution in the presence of a finite strain field.    Expanding the differential susceptibility as a series in  $\varepsilon_{res}$ gives:
\begin{equation}
   \left[\frac{\partial \phi}{\partial \varepsilon}\right]_{\varepsilon = \varepsilon_{res}} = \frac{C}{T-T_0} - \frac{C_2^3\varepsilon_{res}^2}{(T-T_0)^3}. 
\end{equation}
The solid line in Fig. \ref{fig:chi}(b) is a fit to this expression with $C = (1.93\pm 0.24) \times 10^3$ K and $C_2 = (5.3 \pm 3.1) \times 10^3$ K and $T_0 = 116$ K. The dotted line shows the result with $C_2=0$, which should be the response in the absence of $\varepsilon_{res}$.  The value of $C$ is approximately 40\% of that reported previously \cite{EFGnematicity}. The reason for reduction could be errors in the strain-to-voltage calibration factor, $\alpha$. In previous studies using a strain gauge to estimate $\alpha$, this quantity was observed to decrease by up to 8\% at frequencies up to 10 kHz \cite{Hristov2018}, therefore it is possible that the strain levels reported here are overestimated, and the magnitude of the nematic susceptibility is underestimated.

\section{Discussion}

\begin{figure}[!h]
    \includegraphics[width=\linewidth]{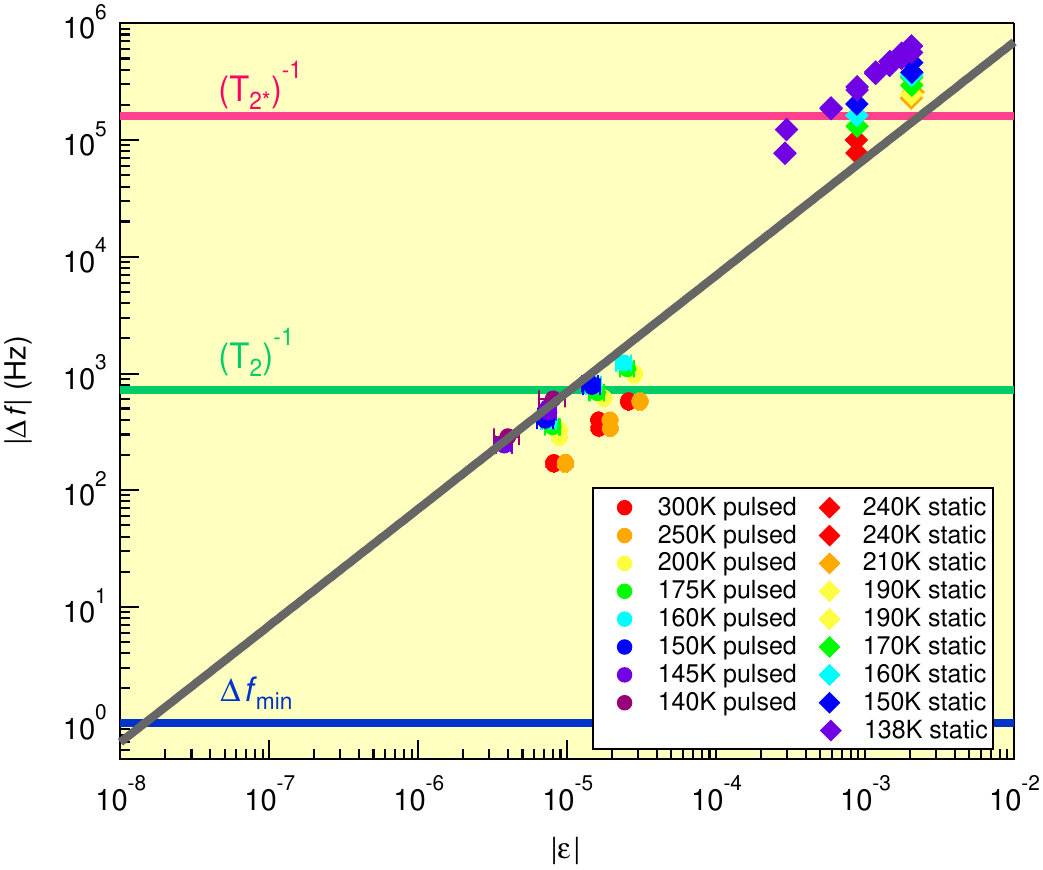}
    \caption{\label{fig:loglogplot}  Comparison of the  frequency offset, $|\Delta f|$, versus strain, $\varepsilon$, measured by static ($\blacklozenge$, reproduced from \cite{EFGnematicity}) and pulsed ($\bullet$) techniques. The pink and green lines correspond to $T_{2*}^{-1}$ and $T_2^{-1}$ for BaFe$_2$As$_2$, respectively, and the blue line corresponds to the minimum detection level, as discussed in the text. }
\end{figure}

Figure \ref{fig:loglogplot} compares the NMR frequency shifts measured with pulsed strain and static strain. Despite several orders of magnitude difference in the strain, the slopes are approximately the same for the different temperatures. This result confirms that our approach using the calibration factor $\alpha$ for pulsed strain  captures the nematic susceptibility reasonably well, despite utilizing strain levels that are 2-3 orders of magnitude smaller.  

The  green and red horizontal lines in Fig. \ref{fig:loglogplot} indicate the homogeneous and inhomogeneous linewidths in this material, which set the scale for the sensitivity limits for the measured frequency shift, $\Delta f$.  It is clear that the pulsed strain approach enables measurements of the nematic susceptibility with much smaller levels of applied strain by taking advantage of the fact that $T_2^{-1}\ll (T_2^*)^{-1}$. For the $^{75}$As, inhomogeneous strain in the crystal due to imperfections or dopants gives rise to inhomogeneous broadening via the nuclear quadrupole interaction.  Although it is possible to detect a frequency shift that is smaller than the linewidth, such an approach requires careful measurements of the full spectrum.  On the other hand, the minimum detectable frequency shift using pulsed strain is determined by the precision of the phase angle of the echo.  The minimum $\Delta f$ we measured was 76 Hz at $\varepsilon = 3.4$ ppm, which required 3.5 hours of measurement time.  The precision of our measurements of $\Delta f$ was $\sim 0.5^{\circ}$ for a spectrum with signal to noise ratio $\sim 5\times 10^{6}$. For $T_2 = 692$ $\mu$s, this corresponds to a minimum detectable frequency shift of $\sim 1$ Hz, shown as the solid blue line in Fig. \ref{fig:loglogplot}. If the frequency shift is entirely due to $\chi_{\eta}$, this corresponds to a sensitivity of $\Delta \eta \sim 1$ ppm.  

The pulsed strain technique enables measurements of nematic susceptibilities via NMR in a broader range of materials than were previously accessible. This is particularly true for materials with inhomogeneous linewidths, such as the hole or electron-doped iron-based superconductors, e.g. 
Ba(Fe$_{1-x}$Co$_x$)$_2$As$_2$.  Although the nematic susceptibility has been measured in the normal state of these systems previously, pulsed-strain NMR measurements can enable measurements of the nematic response below $T_c$ where elastoresistance studies are not feasible. Below $T_c$, the NMR signal is usually reduced in a single crystal because the penetration depth is much smaller than the skin depth, requiring longer signal averaging times. In such cases it is difficult to detect spectral shifts in the presence of static strain, however pulsed strain measurements should be straightforward.   Another advantage of this approach is that it may also be useful for materials that exhibit a large Young's modulus, where DC strain approaches are unable to resolve any linewidth changes.  Alternatively, because the high sensitivity of this approach does not require large strain values in order to determine the response function, it will be useful for measurements of the nematic susceptibility in materials that are particularly fragile.  For example, this technique would be ideal to investigate CaKFe$_4$As$_4$, where $T_c$ is high and the material is nominally pure, yet elastoresistance measurements indicate an enhanced nematic susceptibility in the normal state \cite{Cui2017,Elastoresistance1144}. Observations of the how the nematic susceptibility changes in the superconducting state may shed important light on whether nematic fluctuations play a role in stabilizing the superconducting state \cite{Lederer2017}. More broadly, knowledge of the response of the EFG tensor to strain may provide important insights into the electronic structure calculations  in strongly correlated materials \cite{MeierEFG2002,NQR115electronic}. 

\acknowledgements

We thank A. Chakraborty, J. Barraclough, R. Fernandes, I. Fisher and A. Ward for stimulating discussions. NJC greatly benefited from the International Workshop on the Experimental Advances in the Use of Pressure and Strain to Probe and Control Quantum Matter, PSQM, held in Ames Iowa in May of 2022; many of the key ideas associated with this paper were developed during the workshop time. Work at UC Davis was supported by the NSF under Grants No. DMR-2210613 and PHY-1852581.  Work at the Ames National Laboratory was supported by the U.S. Department of Energy, Office of Science, Basic Energy Sciences, Materials Sciences and Engineering Division.  Ames National Laboratory is operated for the U.S. Department of Energy by Iowa State University under Contract No. DE-AC02-07CH11358.

\section*{References}
\bibliography{pulsedbib}

\end{document}